# Canonical formulation of Thermodynamics, and corresponding wave equation for far from equilibrium case

Adrián Faigón*

Laboratorio de Física de Dispositivos - Departamento de Física - Facultad de Ingeniería-
Universidad de Buenos Aires. Paseo Colon 850 - (1063) Buenos Aires, Argentina

Lagrange and Hamilton equations for thermodynamic evolution near equilibrium as well as Schrodinger-like equation for the non-equilibrium case are obtained extending the CPDQ Principle (Constancy of the product momentum-coordinate uncertainty) to a new conjugate pair where the momentum f is the extension of the mechanical f=p.δq product for the case the coordinate q is not directly observable. Applied to the case of a particle in a one dimension box, the results are in agreement with reported calculated and measured thermal conductance.

--------

*A. Faigón is research fellow at the CONICET (Consejo Nacional de Investigaciones Científicas y Técnológicas). afaigon@fi.uba.ar.

## I. INTRODUCTION

Physics seeks formalization and simplification. The canonical scheme of Mechanics is the paradigm for this aim. There are several proposals for a canonical formulation of thermodynamics, the first attempt due to Helmholtz [1]. His canonical scheme was resumed after the development of quantum mechanics by several researches following DeBroglie effort to identify Helmholtz's monocycles with oscillations of hidden variables [1-3]. Different approaches were more recently developed presenting Hamiltonian formalisms field theory for non-equilibrium thermo-dynamics [4-6].

On the other hand, due, probably, to structural analogies between non-equilibrium thermo-dynamics and quantum evolutions: lack of definite states, uncertainty and statistical interpretation, attempts have been made in order to find Schrodinger like equations for thermodynamic phenomena [7-10].

In this work we resume the line initiated by Helmholtz attempting to develop thermo-dynamics from mechanical considerations. A canonical description of thermodynamic evolutions is derived here from the extension of a new mechanics foundation formalism based in the constancy of the p-δq (momentum-coordinate uncertainty) product (CPDQ) for directly observable degrees of freedom [11]. The extension is based on taking for the coordinate uncertainty the dimension $\Delta q$ of the boundary enclosing the thermodynamic system and hiding the internal degrees of freedom from direct observation. In this case the mechanical f=p·$\Delta q$ product is no longer constant but there is a meaningful new evolutionary coordinate g satisfying f·δg constant. I.e. a new thermodynamical CPDQ principle. From it, and following similar procedures as those used to derive mechanical and quantum mechanical laws, the Lagrangian and Hamiltonian formulation for the thermodynamics of a particle in a box is presented, as well as the Schrodinger like wave equation for the thermal non-equilibrium case. The thermal conductance derived from the presented formulation is in essential agreement with the expressions obtained by completely different ways and experimentally tested [12,13].

## II. INTRODUCING THE EVOLUTIONARY QUANTITIES

**Review of main mechanical and thermo-diynamical quantities**

In a previous work we have shown that mechanics can be founded on a principle (Constancy of P deltaQ product or CPDQ) stating the constancy of the product

$$f = p\delta q \qquad (1)$$

being p the q-conjugated momentum, and δq the minimum uncertainty with which q --an observable coordinate-- can be determined [11].

It was also shown that thermodynamics appears as a straightforward extension or this formulation for the case the particles are hidden from direct observation. The position uncertainty is determined by the enclosing boundaries and is, therefore, disengaged from the momentum p, allowing the extended definition of the f-product

$$f \equiv p\Delta q \qquad (2)$$

to vary. Thermodynamical magnitudes were identified as follows:
Temperature T by

$$kT \equiv p\dot{q} \qquad (3)$$

with k the Boltzman constant; and entropy S through

$$\frac{dS}{k} \equiv d\ln f. \qquad (4)$$

For a system with N non-interacting particles in equilibrium, sharing the same $\Delta q$, $p \cdot dq/dt$ and $\ln f$ are mean values; and k is replaced by Nk in (4).

Completing this partial review, the change in what was termed the extended Hamiltonian in [11], which was found to be the "reversible heat" Qr is

$$dH^e \equiv dQr = TdS = p\dot{q}\frac{df}{f} = \frac{|\dot{q}|}{\Delta q}df \qquad (5)$$

**The new coordinate "number of interaction" or "advance"**

Let us now define the new coordinate g by

$$dg \equiv \frac{|dq|}{\Delta q} \qquad (6)$$

g is an ever increasing number which measures the number of uncertainty intervals traveled by the particle. $|dq/dt|/\Delta q$ is, therefore, the frequency dg/dt of (potential) interactions of the particle with its surroundings for degree of freedom, and eq. (5) may be rewritten

$$dH^e(f) = \dot{g}\,df \qquad (7)$$

**The evolutionary energy**

Comparing last expression (7) for the hamiltonian with its expression in terms of mechanical variables Ref [11] Eq.(22),

$$dH(q,p) = \dot{q}\,dp + p\dot{q}\frac{d\ln\delta q}{dq}dq, \qquad (8)$$

we find that the r.h.s in (7) is formally equivalent to the kinetic energy in (8), i.e. the product of a "velocity" dg/dt, times a change in "momentum" f. This quantity, with energy units, changing its value with the change of the momentum f will be called "evolutionary energy"

$$dKev \equiv \dot{g}\,df$$

for last expression distinguishes evolutionary systems (dQr≠0) from conservative ones (adiabatic and reversible, df=0 <==> dH≡0).

**Evolutionary force and potential**

The equivalent to the energy-work theorem is, in terms of the new variables g and f,

$$\dot{g}\,df = \frac{dg}{dt}\dot{f}\,dt = \dot{f}\,dg \qquad (9)$$

The r.h.s. is the product of a time derivative of momentum f times a "displacement" dg, i.e. evolutionary work.
It can be shown by direct replacement from the definitions that the analogy is extended to

$$\dot{f} = -f\dot{g}\cdot\frac{d\ln\delta g}{dg} \qquad (10)$$

if $\delta g$ is, according to the definition of dg in Eq. (6),

$$\delta g = \delta q/\Delta q \quad , \qquad (11)$$

the fraction of the actual uncertainty $\Delta q$, occupied by the mechanical uncertainty $\delta q$ (that corresponding to a freely observable particle). $\delta g$ is, therefore, the uncertainty in the number of interaction g and its value is bound between 0+ and 1. Eq. (10) defines the "evolutionary force" (time derivative of the ev-momentum) as the g-derivative of the ev-work, or, in the case of "conservative ev-forces", as the gradient of the ev-potential

$$dVev(g) \equiv f\dot{g}\cdot d\ln\delta g(g) \qquad (12)$$

## III. CANONICAL FORMULATION OF THERMODYNAMICS

**The CPDQ principle in Thermodynamics**

Thermodynamics can be founded on a principle stating the constancy of the product of momentum f times coordinate uncertainty $\delta g$

$$F \equiv f \cdot \delta g = const \qquad . \qquad (13)$$

The principle has been called in Ref [11] the Constancy of P-deltaQ product, CPDQ, and we retain this denomination here. Being the values of $\delta g$ bounded between 0 and 1, the minimum value for f is that of the constant. The particular case $\delta g=1$ corresponds to the mechanical case, indicating that the constant in (13) is $\hbar/2$; other constant values for $\delta g$ correspond to adiabatic processes.

**Temporal evolution. Lagrange equation**

If $\delta g$ is not a constant, the evolution of the system is found through the time derivative of eq. (13)

$$\dot{F} = \dot{f} \cdot \delta g + f \cdot (\dot{\delta g}) = 0. \qquad (14)$$

Being $\delta g(t)$ a variation of the "evolutionary trajectory" g(t) as in Fig.1,

$$(\dot{\delta g}) = \frac{d(\delta g)}{dt} = \frac{d(g+\delta g)}{dt} - \frac{dg}{dt} = \delta \dot{g}, \quad (15)$$

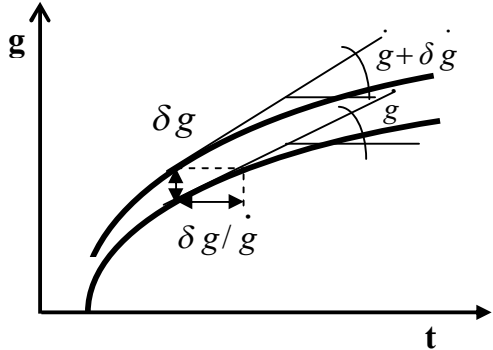

Fig.1 **The actual g(t) and varied (g+$\delta$g)(t) ev-trajectories**

the r.h.s. being the uncertainty variation of the velocity dg/dt, i.e. the change in dg/dt between two "mechanically indistinguishable" trajectories. "Mechanically" because in one $\delta g$, the particle moves one $\delta q$ which is the limit of the mechanically distinguishable.

Using Eq. (15), the middle member of Eq. (14) expresses the variation of a function $L_{ev}(g,dg/dt)$ across the bundle of undistinguishable trajectories, what has been called the uncertainty variation,

$$\delta L_{ev}(g,\dot{g}) = \dot{f} \cdot \delta g + f \cdot \delta \dot{g}, \quad (16)$$

from which a Lagrange like equation,

$$\frac{d}{dt}(\frac{\partial L_{ev}}{\partial \dot{g}}) - \frac{\partial L_{ev}}{\partial g} = 0 \qquad (17)$$

results for g(t). The solution together with the momentum-velocity relationship, implicit in the Lagrangian function, ($f=\partial L_{ev}/\partial(dg/dt)$), gives the system evolution.

**Hamilton formalism**

Complying with the least observables changes prescribed by CPDQ, the time derivative of whichever quantity X, is, in this context,

$$\dot{X} = \frac{\dot{g}}{\delta g} \cdot \bar{\delta} X, \qquad (18)$$

where $\bar{\delta} X$ is the change in X when the system "advances" one $\delta g$. Using (18) in Eq. (16) gives

$$\dot{F} = \dot{g} \bar{\delta} f + f \cdot \frac{\dot{g}}{\delta g} \bar{\delta} \delta g \qquad (19)$$

which is the change corresponding to an *advance* $\delta g$ along the "evolutionary trajectory" g(t) of a function H(f,$\delta$g) satisfying

$$\dot{g} = \frac{\partial H}{\partial f} \quad and \quad f\frac{\dot{g}}{\delta g} = \frac{\partial H}{\partial \delta g} \quad (20a,b)$$

Eq. (20a) is one of Hamilton's canonical equations. Eq. (20b) may be modified rewriting Eq. (19)

$$\dot{F} = \dot{g} \bar{\delta} f + f \frac{\dot{g}}{\delta g} \frac{\partial \delta g}{\partial g} \bar{\delta} g \equiv \bar{\delta} H_{ev} \qquad (21)$$

which expresses, for the case $\delta g(g)$ --or equivalently the potential function-- is known, the *uncertainty* change $\bar{\delta} H_{ev}$ of a function $H_{ev}$(f,g). The ev-Hamiltonian function defined in (21) therefore satisfies

$$dH_{ev} = dK_{ev} + dV_{ev}$$

and gives, instead of Eq. (8b),

$$f \dot{g} \cdot \frac{\partial \ln \delta g}{\partial g} = \frac{\partial H_{ev}}{\partial g} \qquad . \qquad (22)$$

On the other side, replacing in Eq. (9) dg/dt.$\bar{\delta}$f by its original form dfdt.$\delta$g, using CPDQ dF/dt=0, and $\bar{\delta}$g=$\delta$g, results in

$$\dot{f} = -f\dot{g}\frac{\partial \ln \delta g}{\partial g} \quad , \quad (23)$$

which is the already obtained Eq. (10) for the evolutionary force. From Eqs. (22) and (23),

$$\dot{f} = -\frac{\partial H_{ev}}{\partial g}, \quad (24)$$

the second Hamilton's canonical equation.

**Ev-Hamiltonian and Lagrangian in TD quantities**

Let us identify the components of the ev-Hamiltonian and Lagrangian

$$dH_{ev}(g,f) = \dot{g}\,df + f\dot{g}\frac{\partial \ln \delta g}{\partial g}dg \quad (25)$$

and

$$dL_{ev}(g,\dot{g}) = f.d\dot{g} - f\dot{g}\frac{\partial \ln \delta g}{\partial g}dg, \quad (26)$$

in usual thermodynamic quantities. The "kinetic" term of the ev-Hamiltonian is, as shown, the reversible heat absorbed by the system, or TdS as results from multiplying and dividing it by f, and noting that $f\dot{g} = p\dot{q} = kT$. The potential part is minus the same quantity, for CPDQ imposes $dH_{ev}=0$, expressed through the changes in $\delta g$ as a result of changes in "position" g,

$$dH_{ev}(f,g) = TdS(f) + kTd\ln \delta g(g)$$
$$= TdS(f) - dQrev(g) \quad (27)$$

Regarding the Lagrangian, the potential term is the same, but the "kinetic" term is

$$dL_{ev}(\dot{g}) = f.d\dot{g} = d(f\dot{g}) - \dot{g}\,df$$
$$= kdT(f,\dot{g}) - TdS(f) \quad (28)$$

Thus, the Lagrangian in thermodynamic quantities is

$$dL_{ev}(g,\dot{g}) = kdT - TdS + dQrev(g), (29)$$

and, for completeness

$$dH_{ev} + dL_{ev} = kdT. \quad (30)$$

**Example. Heat transmission at constant volume**

Consider a system consisting of a particle in a one-dimension box. At constant volume ($\Delta q$), the momentum-velocity relationship is, from definitions,

$$f(\dot{g}) = m.\Delta q^2\, \dot{g}. \quad (31)$$

Let us show the Lagrangian

$$L_{ev}(g,\dot{g}) = \frac{1}{2}m.\Delta q^2\, \dot{g}^2 - \frac{k}{2}(Te - T_0)\mathrm{e}^{(g_0-g)} \quad (32)$$

describes correctly the evolution for the system starting at temperature $T_0$ immersed at $g=g_0$ in a thermal bath at temperature Te. The Euler-Lagrange equation for g writes

$$m.\Delta q^2\, \ddot{g} - \frac{k}{2}(Te - T_0)\mathrm{e}^{(g_0-g)} = 0. \quad (33)$$

We use

$$\ddot{g} = \frac{d\dot{g}}{dt} = \frac{d\dot{g}}{dg/\dot{g}} = \frac{d(1/2.\dot{g}^2)}{dg} \quad (34)$$

to rewrite (33)

$$d(1/2.\dot{g}^2) = \frac{k(Te - T_0)}{2m.\Delta q^2}\mathrm{e}^{(g_0-g)}\,dg \quad (35)$$

which integrates to

$$\dot{g}^2 - \dot{g}_0^2 = \frac{k(Te - T_0)}{m.\Delta q^2}\left[1 - \mathrm{e}^{(g_0-g)}\right]. \quad (36)$$

The complete solution g(t) is

$$g(t) = \ln\frac{[(\dot{g}_e^2 - \dot{g}_0^2)\mathrm{e}^{-\frac{\dot{g}_e t}{2}} + (\dot{g}_0 + \dot{g}_e)^2\,\mathrm{e}^{\frac{\dot{g}_e t}{2}}]^2}{4\dot{g}_e^2\,(\dot{g}_0 + \dot{g}_e)^2} \quad (37)$$

where $\dot{g}_o$ and $\dot{g}_e$ are the ev-velocities corresponding to initial and equilibrium temperatures. The ev-trajectories are straight lines at fixed temperatures as shown in Fig. 2 out of the transition region beginning at t=0.

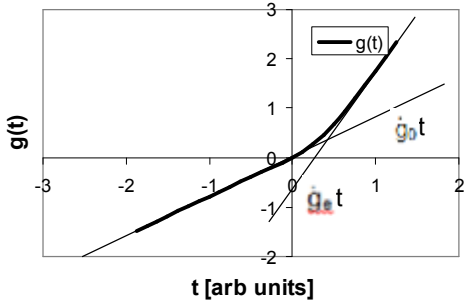

**Fig. 2** The ev-trajectory g(t) for a transition from a temperature To to Te=10.To at t=0.

The Thermodynamic interpretation of the solution is straightforward. Using (31), and the identity $f\dot{g} = kT$, (36) rewrites

$$T - T_0 = (Te - T_0)\left[1 - e^{(g_0 - g)}\right]. \qquad (38)$$

The conductance involved in the process is $\sigma \equiv 1/(Te-T).dQ/dt$, where $dQ = 1/2.CvdT$ --the 1/2 is to take the amount of heat interchanged through one wall--, $dt = dg/(dg/dt)$ and T from (38), gives

$$\sigma = \frac{Cv\dot{g}}{2} = \frac{CvkT}{2f} \leq \frac{\pi k^2 T}{h} \qquad . \qquad (39)$$

The r.h.s expresion is the the maximum value for the conductance, corresponding to Cv=k/2 for one degree of freedom, and the minimum value for f, f=ℏ/2. It has the same form and comparable value (differing in a factor π/3) to the quantum heat conductance predicted by Roukes and measured by Rego et al. [12,13].

The force responsible for the change in momentum f and for deviating the trajectory g(t) from the straight-line it follows in equilibrium, is, from the second term of the r.h.s. in (32) representing the potential energy, and (38),

$$evForce = \frac{k}{2}(Te - T) \qquad (40)$$

From last result and (4) the rate of change for the entropy is

$$\dot{S} = k\frac{\dot{f}}{f} = \frac{k^2(Te-T)}{2f} = \frac{k^2(Te-T)}{2\Delta q\sqrt{mkT}} \qquad . \qquad (41a)$$

In the ev-representation $(g, \dot{g})$, $\dot{S}$ is, straightforward from the Lagrangian,

$$\dot{S} = -k.\frac{\partial L_{ev}}{\partial g} \bigg/ \frac{\partial L_{ev}}{\partial \dot{g}} \qquad . \qquad (41b)$$

## IV. FAR FROM EQUILIBRIUM AND WAVE FUNCTION

The power of the CPDQ formulation resides in its common root for a trajectory description (classical) and a wave description (quantum). The conditions for one or the other description to hold are detailed in Ref [11] together with the way for obtaining the corresponding equations.

Having presented a CPDQ formulation for Thermodynamics, and its Lagrangian and Hamiltonian trajectory equations, the next step is to show in which case and how a wave description can be made. We will follow in a shorter and slightly modified way the reasoning given in detail in Ref [11]:

If the condition for trajectory description $\left|\overline{\delta}\delta g\right| << \delta g$ is not fulfilled, then δg(t) is not well defined. An alternative representation for δg is suitably given by the generalized Cramer-Rao (CR) inequality [11,14]

$$(\delta g)^2 \int 4\psi'^*\psi' dg \geq 1 \qquad (42)$$

where $\psi^*(g).\psi(g) = P(g)$ is the probability density function for the observable g, the prime denotes the g derivative. Using CPDQ eq. (13), taking mean value for $f^2$, which is affected from the same lack of certainty, and considering the limiting case for the CR inequality,

$$\hbar^2 \int \frac{P}{\psi^*\psi} \psi'^*\psi' dg = \int P f^2 dg \qquad (43)$$

which is true if

$$-i\hbar\frac{\partial \psi}{\partial g} = f\psi \qquad . \qquad (44)$$

Last expression defines the momentum operator. It serves to construct the ev-Hamiltonian operator from its "classical" expression in (25), and the corresponding wave equation, proceeding as in mechanics

$$\hat{H}ev(\psi) \equiv$$
$$\hat{T}ev(-i\hbar\frac{\partial \psi}{\partial g}) + Vev(g)\psi = Qrev_{eq}\psi \qquad .(45)$$

In the case of the above example, the equation for the far from equilibrium thermal flow at constant volume is - using (32) and (30)-

$$\frac{\hbar^2}{2m\Delta q^2}\frac{\partial^2 \psi}{\partial g^2} + \frac{k}{2}To\psi = 0 \qquad (46a)$$

for g<go in which the system is put in thermal contact with the reservoir, and

$$-\frac{\hbar^2}{2m\Delta q^2}\frac{\partial^2 \psi}{\partial g^2} + [\frac{k}{2}(Teq-To)e^{(go-g)} - \\ -\frac{k}{2}Teq]\psi = 0 \qquad (46b)$$

for g≥go.

The solutions, shown in Fig. 3 and detailed in the Appendix, are, out of the transition region, harmonic waves with wave numbers

$$kev = \sqrt{\frac{kTo}{2Eoo}} \text{ and } kev = \sqrt{\frac{kTeq}{2Eoo}} \quad (47 \text{ a,b})$$

for the initial and final states, being Eoo the energy of the fundamental state of the particle in the box (δq = Δq) Eoo = 1/ 2m.(ℏ/Δq)².

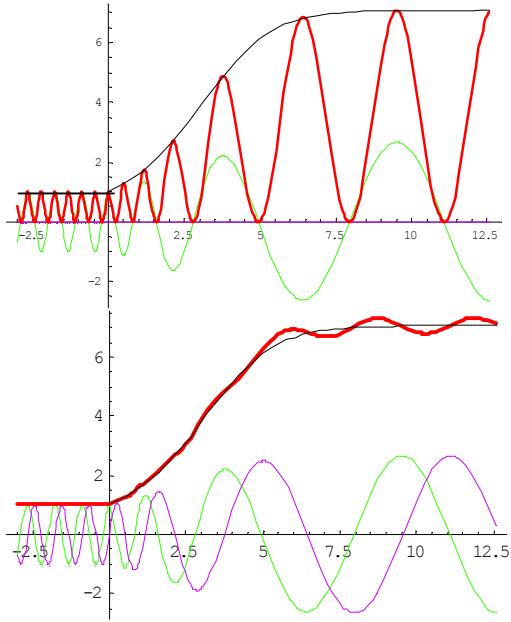

**Fig. 3.** Two different ev-Wave functions, solutions of eq. (46) corresponding to different initial conditions for a one particle - one dimension system evolving between two arbitrary temperatures (from 50 to 1 times the fundamental state). At g=0, the system hitherto in equilibrium get in thermal contact with the low temperature reservoir.
**In violet (coincident with the x axis in the upper plot) the Imaginary part of the wave functions, in green the Real part, in red is |ψ|2 and in black is (To/T)$^{1/2}$.**

In order to check the solution, the mean values of the temperature over half periods were calculated with the obtained wave function

$$<T> = \frac{\hbar^2}{km\Delta q^2}\frac{\int \psi^* \frac{\partial^2 \psi}{\partial g^2} dg}{\int \psi^* \psi \, dg} \qquad (48)$$

The results are plotted together with those obtained from the "classical" treatment given by eq. (38) for the transition from the fundamental state to a temperature Teq 50 times higher.

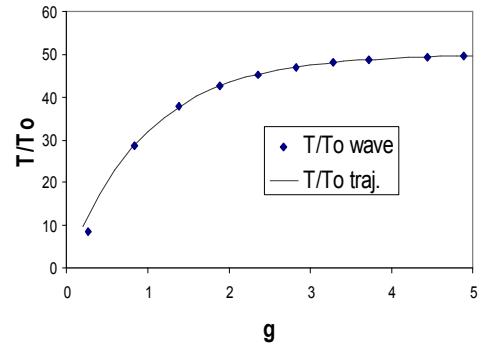

**Fig. 4. Trajectory and wave computation -eqs. (37) and (48)- of the temperature evolution To=1 to Te=50.**

The mean values over half period intervals were assigned in the g-axis of Fig. 4 to the middle point of the interval. This is a cause of discrepancies as the temperature changes --classically-- more in the first half of the interval than in the second. Even though, the wave representation expected value coincides within the plot resolution with the calculated "classical" one with the exception of the first point for which the misbalancing between first and second half of the interval is more evident.
The amplitude of the wave function was found to evolve as $T^{-1/4}$, as shown in Fig. 3, meaning that the probability ψ*ψdg of finding the particle in the interval dg around g goes as $g^{-1}$dg, i.e. proportional to the time spent by the particle in the interval dg.

## V. DISCUSSION

**Statistical interpretation**

The above results correspond, such as they were presented, to one particle in a one dimensional box. Put this system in thermal contact with a thermostat, it is expected to observe a distribution of mechanical states. As f was defined through mechanical quantities, f≡p.Δq, the question arises about the meaning of the obtained results. We want to show here that the same results admit a statistical interpretation.

Were the wave function ψ representing mechanically defined microstates, an adequate representation for the one particle system in thermal contact with a thermostat at temperature T would be given by a Gibbs distribution of states, i.e. a probability density

$$P = \sum P_n \psi_n^* \psi_n \quad . \qquad (49)$$

where the $\psi_n$'s are a set of eigenfunctions with eigenvalues En of the corresponding SE, and Pn given by

$$P_n = \frac{1}{z} e^{-\frac{En}{kT}} \qquad (50)$$

with z the states sum. Such a representation was introduced in [15].
Two ways in which the quantities involved in the formulation (f,g) can be read as statistical results, i,e. representing thermodynamic and not mechanical states, are as follows.
It has been shown [16] that the probability density P=|ψ|² resulting from the solution of the Schrodinger equation for one particle takes, in the classical limit where the WKB approximation is valid, the same form as a classical probability distribution function for an ensemble of particles. As expected in a classical statistical ensemble, the probability density is proportional to the inverse of the classical velocity; and the mean velocity, given by the current density, is in this case equal to the classical velocity. Translated to our thermodynamical treatment for a particle in a box, it means that the probability density associated to the obtained thermal wave function, corresponds, not far from equilibrium, to a distribution of particles with mean (r.m.s.) "velocity" $\dot{g}(g) = 1/\Delta q.(kT/m)^{1/2}$; and is, then, proportional to $T^{-1/2}$ as shown in Fig. 3, so Pdg is proportional to the time spent in dg as classically expected.

A broader statistical interpretation follows the observation that the momentum f is proportional to the number of available states Ω for the system with given energy kT, and, in this case, to z, the canonical partition function for the one degree of freedom system here considered:

$$f \equiv p\Delta q = (mkT)^{1/2} V^{1/3} \qquad (51)$$

and

$$z \equiv \sum e^{-\frac{En}{kT}} = \frac{1}{h}(2\pi mkT)^{1/2} V^{1/3}, \qquad (52)$$

therefore

$$f = (2\pi)^{1/2} \hbar z \qquad . \qquad (53)$$

But expression (52) for z is in turn, the partition function per degree of freedom for a system of N non-interacting particles in a volume V, $z=(N!\ Z)^{1/3N}$, where Z is the partition function for the system.

For f in (53) to be interpreted in such a way it suffices to use for p in the definition of f, its r.m.s. value over the system, consistent with $kT = <p.\dot{q}>$ as stated in section II. In this way the momentum f whose evolution is obtained through the equations of Lagrange, or Schrodinger, for close-to or far-from equilibrium respectively, is a statistic which in equilibrium coincides (constants apart) with the partition function per degree of freedom of a non interacting particles system.

**Heisenberg uncertainty relationship and fluctuations**

Both treatments given above, the trajectory representation corresponding to close to equilibrium evolutions and the more general wave representation necessary for non equilibrium evolutions, are developed from the general CPDQ principle, stating, in the thermodynamical version, that the evolution is characterized by f, or equivalently by δg. Now, given f and the f-$\dot{g}$ relationship --via the Lagrangian function--, the temperature is also completely defined. But the homomorphism with the mechanical treatment, emphasized along all the work, reminds us that uncertainty-fluctuations on the value of the well defined momentum will become apparent in observations. This stems from the Heisenberg uncertainty relationship. As a consequence we are faced --as discussed in [11]- with f.δg=ℏ/2 wich complety determines f, together with Δf.Δg≥ℏ/2, meaning that the f assigned to the interval δg=1/2.$k_{ev}$=ℏ/2f --see appendix in [11]-- has an uncertainty Δf whose lower limit is precisely f. It means for our one particle system in thermal contact with a thermostat at temperature T, to exhibit energy fluctuations of magnitude kT as expected.

**Helmholtz Canonical Scheme**

Helmholtz proposed a canonical scheme for thermodynamics lying on the assumption of a formal cyclical variable ε whose time derivative is the temperature [1,2]

$$T = \dot{\varepsilon} \qquad . \qquad (54)$$

If E is the generalized force associated to the coordinate ε, then

$$dU = Ed\varepsilon - PdV = E\dot{\varepsilon}\,dt - PdV \quad (55)$$

which compared with the fundamental equation

$$TdS = dU + PdV$$

yields

$$E = \dot{S} \quad (56)$$

allowing a Lagrangian $L = L(V, \dot{V}, \dot{\varepsilon})$ to describe the system in equilibrium with the entropy S as the conjugate momentum of ε.

The Helmholtz canonical scheme is reproduced from present results by taking for its elusive (formal) variable dε=da/k, with

$$da \equiv fdg \quad (57)$$

satisfying (54), and its conjugated momentum

$$ds \equiv df/f \quad (58)$$

i.e. the entropy in k units. The product

$$da.ds \equiv dg.df \quad (59)$$

preserves the phase space volume granting the canonicity of the transformation.
Lagrangian and Hamiltonian write

$$dL(a,\dot{a}) = s(\dot{a})d\,\dot{a} + \frac{\partial Qrev}{\partial a}da \quad (60)$$

and

$$dH(a,s) = \dot{a}\,ds - \frac{\partial Qrev}{\partial a}da \quad (61)$$

Giving for the evolutive generalized force associated to the new variable a

$$-\frac{\partial Qrev}{\partial a} = -\frac{1}{f}\frac{\partial Qrev}{\partial g} = \frac{\dot{f}}{f} = \dot{s}, \quad (62)$$

as in (56); and, for completeness

$$dL(a,\dot{a}) + dH(a,s) = d(\dot{a}.s) = d(TS). \quad (63)$$

In addition to satisfy Helmholtz's requirements, the presented canonical scheme is formulated in terms of the relativistic invariant magnitudes entropy and action as the new variable a is, from definitions

$$a = \int f dg = \int p dq = \text{Maupertius action}.$$

In the light of these results, our variable g

$$g = \int \frac{1}{f}da$$

is the Maupertius action measured in f cycles; i.e. the phase=Action/ℏ in the classical limit of quantum mechanics in which f = ℏ/2.

**Appendix A**

Solution of eqs. (46) is for g<go

$$\psi(g) = Ae^{\sqrt{a-c}.g} + Be^{-\sqrt{a-c}.g} \quad \text{(A1a)}$$

and, for g>go

$$\psi(g) = C(-1)^{-i\sqrt{c}}\Gamma(1-2i\sqrt{c})I_{-2i\sqrt{c}}(2\sqrt{a.e^{-g}}) + \\ + D(-1)^{i\sqrt{c}}\Gamma(1+2i\sqrt{c})I_{2i\sqrt{c}}(2\sqrt{a.e^{-g}})$$
(A1b)

where Γ is the gamma function, $I_r$ is the modified Bessel function of the first kind,

$$a = \frac{k(Teq - To)}{2E_{oo}}, \text{ and } c = \frac{kTeq}{2E_{oo}}.$$

The constants A to D were chosen to satisfy boundary conditions (continuity of the function and its first derivative) and normalization.

**References**

1. -J. Andrade e Silva, "On Another Formulation of the de Broglie's Hidden Thermodynamics" International Journal of Theoretical Physics, **3-1**, pp. 67-76 (1970),
2. -L. DeBroglie, La Thermodinamique de la Particle Isolée, Gauthier-Villars Ed., Paris, 1964
3. -M. Campisi, Studies in History and Phylosopy of Modern Physics 36, 275-290 (2005)
4. -K. Gambár, "Hamilton-Lagrange formalism of nonequilibrium thermodynamics", Phys. Rev E, 50, 1227, (1994).
5. -M. Grmela, "Hamiltonian extended thermodynamics", J. Phys. A. 23, 3341 (1990).
6. -S. Sieniutycz, "Canonical formalism, fundamental equation, and generalized thermomechanics for irreversible fluids with heat transfer, Phys. Rev. E, 47, 1765, (1993)
7. -M. Mehrafarin, " Canonical operator formulation of nonequilibrium thermodynamics", J. Phys. A: Math. Gen. 16, 5351, (1993)
8. -M. Kostin, "Thermal Schrodinger equation", Phys. Rev. E, 52-1, 229, (1995)
9. -B. R. Frieden, "Schrodinger link between nonequilibrium thermodynamics and Fisher information", Phys. Rev. E 66, 046128-1 (2002)
10. -Badiali, "Entropy, time-irreversibility and the Schrödinger equation in a primarily discrete spacetime", J. Phys. A: Math. Gen. 38, 2835 (2005)
11. -A. Faigon, " Uncertainty and information in classical mechanics formulation. Common ground for thermodynamics and quantum mechanics", accompanying article
12. -Roukes, ,Nature (London), 404, 974 (2000).
13. -L. Rego, " Quantized Thermal Conductance of Dielectric Quantum Wires", Phys. Rev. Lett. 81-1, 232 (1998).
14. -M. J. W. Hall, "Quantum properties of classical Fisher information", Phys. Rev. A 62, 012107-1 (2000) .
15. -J. C. Slater, Phys. Rev. 38, 237 (1931)
16. -D. Bohm, "Quantum Theory", 1951, , Dover N.Y. facs. Ed. 1989, Ch 12 p. 268